\pdfoutput=1

\documentclass[published]{JHEP3} 

\JHEP{00(0000)000}

\JHEPspecialurl{http://jhep.sissa.it/JOURNAL/JHEP3.tar.gz}

\usepackage{epsfig,multicol,bbm}

\newcommand\fverb{\setbox\pippobox=\hbox\bgroup\verb}
\newcommand\fverbdo{\egroup\medskip\noindent%
			\fbox{\unhbox\pippobox}\ }
\newcommand\fverbit{\egroup\item[\fbox{\unhbox\pippobox}]}
\newbox\pippobox

\def\gtap{\mathrel{ \rlap{\raise 0.511ex \hbox{$>$}}{\lower 0.511ex
   \hbox{$\sim$}}}}
\def\ltap{\mathrel{ \rlap{\raise 0.511ex
    \hbox{$<$}}{\lower 0.511ex \hbox{$\sim$}}}}
\newcommand{\bea}{\begin{eqnarray}}
\newcommand{\eea}{\end{eqnarray}}
\def\beq{\begin{equation}}
\def\enq{\end{equation}}
\def\ba{\begin{eqnarray}}
\def\ea{\end{eqnarray}}

\newcommand{\g}{\gamma}

\newcommand{\be}{\begin{equation}}
\newcommand{\ee}{\end{equation}}
\newcommand{\gag}{g_{a\gamma}}
\def\<{\langle}
\def\>{\rangle}

\title{Hints of an axion-like particle mixing in the GeV gamma-ray
  blazar data?}

\author{ {Olga Mena$^{a}$ and Soebur Razzaque$^{b}$ \\ {$^{a}$IFIC,
      Universidad de Valencia-CSIC, E-46071, Valencia, Spain}
    \\ {$^{b}$Department of Physics, University of Johannesburg, PO
      Box 524, Auckland Park 2006, South Africa} \\ E-mail:
    \email{omena@ific.uv.es}, \email{srazzaque@uj.ac.za}}}

\received{January 1, 0000} 		
\revised{January 1, 0000}
\accepted{January 1, 0000}		

\preprint{\hepth{9912999}}	

\abstract{Axion-Like Particles (ALPs), if exist in nature, are
  expected to mix with photons in the presence of an external magnetic
  field.  The energy range of photons which undergo strong mixing with
  ALPs depends on the ALP mass, on its coupling with photons as well
  as on the external magnetic field and particle density
  configurations.  Recent observations of blazars by the {\em Fermi
  Gamma-Ray Space Telescope} in the 0.1--300 GeV energy range show a
  break in their spectra in the 1--10 GeV range.  We have modeled this
  spectral feature for the flat-spectrum radio quasar 3C454.3 during
  its November 2010 outburst, assuming that a significant fraction of
  the gamma rays convert to ALPs in the large scale jet of this
  blazar.  Using theoretically motivated models for the magnetic field
  and particle density configurations in the kiloparsec scale jet,
  outside the broad-line region, we find an ALP mass $m_a\sim
  (1-3)\cdot 10^{-7}$~eV and coupling $g_{a\g} \sim (1-3)\cdot
  10^{-10}$~GeV$^{-1}$ after performing an illustrative statistical
  analysis of spectral data in four different epochs of emission.  The
  precise values of $m_a$ and $g_{a\g}$ depend weakly on the assumed
  particle density configuration and are consistent with the current
  experimental bounds on these quantities. We apply this method and
  ALP parameters found from fitting 3C454.3 data to another
  flat-spectrum radio quasar PKS1222+216 (4C+21.35) data up to 400
  GeV, as a consistency check, and found good fit.  We find that the
  ALP-photon mixing effect on the GeV spectra may not be washed out
  for any reasonable estimate of the magnetic field in the
  intergalactic media.}


\begin{document}

\section{Introduction}

Axion-photon transition in an external static electric or magnetic
field~\cite{Sikivie:1983ip,Sikivie:1985yu,Raffelt:1987im} is generally
referred to as the Primakoff effect, originally proposed for neutral
pion production by the interaction of a photon with atomic nucleus.
Axion-like particles (ALPs, see the recent review of
Ref.~\cite{Ringwald:2012hr}) , which are a generalization of the QCD
axions~\cite{Peccei:1977hh}, are characterized by these two photon
vertex interactions. Strong theoretical motivation for the existence
of ALPs arises from string theory compactifications, which provide the
so-called \emph{axiverse} scenarios, with plenty of candidates for
ALPs, see
Refs.~\cite{Arvanitaki:2009fg,Acharya:2010zx,Cicoli:2012sz,Ringwald:2012cu}.

The conversion between photons and ALPs is determined by the
ALP-photon coupling $\gag$, and it has been extensively studied in the
context of astrophysical sources in order to search for the
hypothetical ALPs in the optical to X-ray data.  These searches
exclude large regions in the $m_a$--$\gag$ parameter space of the ALP
mass and its coupling to photons (see e.g.\ Ref.~\cite{Arik:2011rx,
Friedland:2012hj} for recent bounds).  In particular, axion energy
losses in stars have provided strong constraints in the ALP parameter
space~\cite{Raffelt:1987yu}.  The emission of ALPs would decrease the
duration of the Helium burning (the Horizontal Branch, HB, stage), and
therefore the number of counts in observations of galactic globular
clusters. The authors of Ref.~\cite{Friedland:2012hj}, using massive
stars, have recently confirmed and sharpened the constrain on the ALP
photon mixing parameter $g_{a\gamma}<(0.8-1) \cdot 10^{-10}$
GeV$^{-1}$.  Future terrestrial based ALP searches, as those from the
ALPS II experiment, are also highly promising~\cite{Bahre:2013ywa}.

Search for ALPs in the $\g$-ray data from astrophysical sources is
being widely discussed in recent years.  Conversion of $\g$ rays to
very light ALPs has been proposed to take place at the sources such as
the Active Galactic Nuclei (AGNs)~\cite{Hooper:2007bq,
  Tavecchio:2012um} and Gamma-Ray Bursts (GRBs)~\cite{Mena:2011xj}; or
in the intergalactic space~\cite{SanchezConde:2009wu, Bassan:2010ya,
  Horns:2012kw}; and in the Milky Way~\cite{Simet:2007sa}.  Depending
on the magnetic field strength of the medium in which the $\g$ rays
propagate, conversion to ALPs could be possible for $m_a \lesssim
10^{-6}$~eV.

In this paper we report on our search for ALPs in the GeV $\g$-ray
data from the best-studied blazar\footnote{A small fraction of AGNs
  with their relativistic jets pointing towards our line of sight.} in
the GeV band, namely the flat-spectrum radio quasar (FSRQ) 3C454.3 at
a redshift $z=0.859$.  {\it Fermi-}Large Area Telescope (LAT)
monitored this blazar daily in its regular survey mode and detected an
extraordinary 5 day outbursts from 2010 November 17th to
21st~\cite{Abdo:2011}.  The spectra of $\g$ rays at different epochs
from 2010 September 1st to December 13th period, which includes the 5
day outbursts, show a break or deviation of the spectra (softening)
from a single power law in the 1--10 GeV range.  We have fitted these
spectra and their breaks using an ALP-photon mixing model for the
$\g$-ray propagation in the magnetized jet that extends to kiloparsecs
($1\, {\rm kpc} = 3\cdot 10^{21}$~cm) outside the blazar's Broad-Line
Region (BLR) at $\lesssim 10^{18}$~cm from the central super-massive
black hole\footnote{This region is filled with dense clouds which emit
  strong atomic transition lines and the distance scale from the
  central black hole depends on the particular line luminosity.},
assuming that the observed $\g$ rays are emitted from the outer edge
of or beyond the BLR.  From our fits we have extracted the values for
$m_a$ and $\gag$, together with the environmental (magnetic field and
particle density) and spectral (power-law index and normalization at
production) parameters that best describe the data.

As a consistency check, we have also fitted {\em Fermi}-LAT and MAGIC
data of another FSRQ PKS1222+216 at redshift $z=0.432$, using the same
$m_a$ and $\gag$ values obtained from fitting 3C454.3 data but varying
the environmental and spectral parameters.  PKS1222+216 data obtained
by {\em Fermi}-LAT \cite{Tanaka:2011gd} and MAGIC
\cite{Aleksic:2011hr} during 2010 June have less constraining power
for the ALP parameters.

The plan of the paper is the following.  We discuss in detail our
ALP-photon mixing model set-up for the blazar jet in Sec.\ 2.  In
Secs.\ 3 and 4 we fit $\g$-ray data of 3C454.3 and PKS1222+216 using
our formalism and report results.  We discuss our results and conclude
in Sec.\ 5.

\section{ALP-photon mixing model for blazar jets}

The $\g$-ray emission region for blazars is a hotly-debated topic.
Observations of rapid variability of fluxes at very-high energy (VHE,
$\gtrsim 100$~GeV) $\g$ rays from several different blazars on time
scales as short as $\sim few$ minutes~\cite{Aharonian:2007ig,
  Albert:2007zd, Aleksic:2011hr} imply that the $\g$-ray emitting
region must be very compact, a size scale of only $\sim 10^{14}$~cm
from the causality condition.  In case the whole jet cross-section is
the $\g$-ray emitting region, then the above size scale also
corresponds to an estimate of the radius from the central
super-massive black hole and is well below the BLR.  On the other
hand, multi-wavelength observations that ``trace'' the propagation of
electromagnetic, from radio to $\g$ rays, emission region in a blazar
jet strongly suggest that the $\g$-ray emission region is at or beyond
the BLR~\cite{Marscher:2008aa}.  Production of $\g$ rays from a large
scale jet also helps to avoid inevitable $\g\g$ pair production by VHE
photons with low-energy photons at small radii.  Rapid variability of
VHE flux in this case is explained as emission from small scale
regions embedded within a large scale jet, i.e., jets within a
jet~\cite{Giannios:2009kh}.  For our modeling we have adopted this
latter scenario of the $\g$-ray production region at a radius at the
outer edge of the BLR or slightly beyond.

We assume that GeV $\g$ rays are emitted from a radius $R\approx
10^{18}$~cm from the central super-massive black hole and propagate
through the kpc scale jet.  The configuration of the magnetic field
and the particle density in the AGN jets are not fully known yet.  It
is expected from the flux-freezing condition that the magnetic field
parallel (poloidal) and perpendicular (toroidal) to the jet velocity
scale with the jet radius, respectively, as $\propto R^{-2}$ and
$\propto R^{-1}$~\cite{Begelman:1984mw}.  Thus, at large radii the
toroidal or transverse component should dominate.  It was also argued
sometime ago that Poynting flux in the AGN jet produce a toroidal
magnetic field $B \simeq 0.4 f^{1/2} (L_w/10^{46}\, {\rm
  erg\,cm}^{-1})(R/10^{18}\, {\rm cm})$~G, which provides magnetic
pressure to confine the BLR clouds~\cite{Rees:1987}.  Here $L_w$ is
the luminosity of a relativistic wind in the jet and $f$ is the
fraction of the wind energy in the Poynting flux.  Reverberation
measurements of the BLR clouds suggest a power-law profile, $R^{-s}$,
of particle density in the jet with $1\lesssim s\lesssim 2$ being
favored~\cite{Kaspi:1999}.  The profile could be steeper (e.g. $s\sim
3$) outside the BLR, which has a particle density $\gtrsim
10^{10}$~cm$^{-3}$~\cite{Kaspi:1999}.

For our ALP-photon mixing model in the jet of the blazar 3C454.3,
motivated by the above discussion, we adopt the following transverse
magnetic field and particle (electron) density profiles
\ba
B_T &=& \phi \left( \frac{R}{10^{18}\, {\rm cm}} \right)^{-1} ~{\rm G},
\nonumber \\
n_e &=& \eta \left( \frac{R}{10^{18}\, {\rm cm}} \right)^{-s} ~{\rm cm}^{-3}.
\label{mag_density}
\ea
We find the normalization parameters $\phi$ and $\eta$ by fitting GeV
$\g$-ray data with our ALP-photon mixing model for different values of
$s = 1$, $2$ and $3$.

To calculate the ALP-photon mixing effect for photons of energy
$\omega$ propagating along the blazar jet, assumed $z$ axis, we
numerically solve the evolution equation~\cite{Bassan:2010ya,
  Mena:2011xj}
\begin{equation} 
i\frac{d}{dz}
\left(\begin{array}{c}A_\perp (z) \\ A_\parallel (z) \\ a (z)
\end{array}\right)=- \left(\begin{array}{ccccccccc}
\Delta_\perp\cos^2\xi+\Delta_\parallel\sin^2 \xi & \cos\xi \sin\xi
(\Delta_\parallel-\Delta_\perp) & \Delta_{a \gamma}\sin \xi \\ \cos\xi
\sin\xi (\Delta_\parallel-\Delta_\perp) &
\Delta_\perp\sin^2\xi+\Delta_\parallel\cos^2 \xi & \Delta_{a
\gamma}\cos \xi \\ \Delta_{a \gamma}\sin \xi & \Delta_{a \gamma}\cos
\xi & \Delta_a \\ \end{array} \right) \left(\begin{array}{c} A_\perp
(z) \\ A_\parallel (z) \\ a (z) \end{array}\right),
\label{eq:evolxi} 
\end{equation} 
with an initial condition $(A_\perp\,, A_\parallel\,, 0)^t = (1/2\,,
1/2\,, 0)$ at $z\equiv R = 10^{18}$~cm, i.e., initially unpolarized
photons.  Here $A_\perp$ and $A_\parallel$ are the electromagnetic
field components, respectively, perpendicular and parallel to $B_T$ in
the $x$-$y$ plane.  The ALP field is denoted with $a$.  $\xi$ is the
angle the transverse magnetic field $B_T$ makes with a fixed $y$ axis
in the $x$-$y$ plane.  For our calculation we fix it to $\pi/4$, and
we will comment on variations of this angle in the following section.
Other different terms in the ALP-photon mixing matrix, with the $B_T$
and $n_e$ given in Eq.~(\ref{mag_density}), are $\Delta_\perp \equiv
2\Delta_{\rm QED} + \Delta_{\rm pl}$, $\Delta_\parallel \equiv
(7/2)\Delta_{\rm QED} + \Delta_{\rm pl}$, and their reference values,
following Refs.~\cite{Bassan:2010ya, Mena:2011xj}, are given as
\ba
\Delta_{\rm QED} &\equiv & \frac{\alpha \omega}{45\pi} 
\left( \frac{B_T}{B_{\rm cr}} \right)^2 
\simeq 1.34\cdot 10^{-18} \phi^2\left( \frac{\omega}{{\rm GeV}} \right)
\left( \frac{R}{10^{18}\,{\rm cm}} \right)^{-2} ~{\rm cm}^{-1},
\nonumber \\ 
\Delta_{\rm pl} &\equiv & -\frac{\omega^2_{\rm pl}}{2\omega} \simeq
- 3.49\cdot 10^{-26} \eta\left( \frac{\omega}{{\rm GeV}} \right)^{-1}
\left( \frac{R}{10^{18}\,{\rm cm}} \right)^{-s} ~{\rm cm}^{-1},
\nonumber \\
\Delta_{a\g} &\equiv & \frac{1}{2}g_{a\g} B_T
\simeq 1.50\cdot 10^{-17} \phi
\left( \frac{g_{a\g}}{10^{-10}\,{\rm GeV}^{-1}} \right)
\left( \frac{R}{10^{18}\,{\rm cm}} \right)^{-1} ~{\rm cm}^{-1},
\nonumber \\ 
\Delta_{a} &\equiv & -\frac{m_a^2}{2\omega}
\simeq -2.53\cdot 10^{-19}
\left( \frac{\omega}{{\rm GeV}} \right)^{-1}
\left( \frac{m_a}{10^{-7}\,{\rm eV}} \right)^2 ~{\rm cm}^{-1}.
\label{matrix_elements}
\ea
Here $\alpha$ is the fine structure constant and $B_{\rm cr} =
4.414\cdot 10^{13}$~G is the critical magnetic field.  The plasma
frequency is defined as $\omega_{\rm pl} = \sqrt{4\pi\alpha n_e/m_e}
=3.713\cdot 10^{-14}\sqrt{n_e/{\rm cm}^{-3}}$~keV.

Although we do not assume a constant magnetic field or particle
density in the ALP-photon mixing region, it is interesting to note
that in such a constant $B_T$ and $n_e$ case the strong mixing of ALPs
with $\gtrsim 1$~GeV photons take place for $B_T \sim
10^{-6}$--$10^{-1}$~G and $n_e\sim 10^7$--$10^8$~cm$^{-3}$ for $m_a
\sim 10^{-7}$~eV and $\gag\sim 10^{-10}$~GeV$^{-1}$ from the low and
high critical-energy conditions $\omega_L \equiv |\omega_{\rm pl}^2 -
m_{a}^2 |/2\gag B_T$ and $\omega_H \equiv 90\pi\gag B_{\rm
  cr}^2/7\alpha B_T$, respectively~\cite{De Angelis:2007yu,
  Bassan:2010ya, Mena:2011xj}.  For our case of varying $B_T$ and
$n_e$ with $R$, however, transitions of photons to ALPs take place
over different radii, $R\sim 10^{18}$--$10^{21}$~cm for $\phi \sim
10^{-3}$ and $\eta\sim 10^9$ in Eq.~(\ref{mag_density}).  We assume a
maximum radius of $R=10^{22}$~cm to solve the evolution equation
[Eq.~(\ref{eq:evolxi})] numerically.  We have checked that the
ALP-photon mixing at larger radii does not contribute to the
0.1--400~GeV energy range of our interest.

\section{Spectral fitting of 3C454.3 data and results}

The 103 day observation, from 2010 September 1st to 2010 December
13th, of the blazar 3C454.3 by the {\em Fermi}-LAT~\cite{Abdo:2011}
constitutes of 4 epochs: (i) An initial quiet or pre-flare period;
(ii) A 13 day long plateau period; (iii) 5 day outburst or flare; and
(iv) A post-flare period.  Data points and upper limits of the
$\g$-ray energy spectra ($\nu F_\nu \equiv E^2\,dN/dE$) in these 4
epochs are shown in Fig.~\ref{fig:spectralfits}.  To fit these
data\footnote{We do not fit the last data points in the flare or
  plateau epochs as they are absent in the quiet and post-flare
  epochs.} with our ALP-photon mixing model, we assume an intrinsic
single power-law spectrum for $\g$ rays, $\propto E^{-\Gamma}$, at the
production region.  This spectrum is modified by a normalized
suppression factor defined as
\be
S(E) = 2\left[ |A_\parallel (E) |^2 + |A_\perp (E)|^2 \right],
\label{suppression}
\ee
where $A_\parallel (E)$ and $A_\perp (E)$ are the solutions of
Eq.~(\ref{eq:evolxi}) with $\omega \equiv E (1+z)$ understood.  The
final observed energy spectrum is then
\be
E^2\, dN/dE = C E^{-\Gamma+2} S(E),
\label{spectrum} 
\ee
where $C$ is measured in erg~cm$^{-2}$~s$^{-1}$.

Our ALP-photon mixing model for blazar jet has six free parameters:
the normalizations for the jet magnetic field ($\phi$) and electron
density ($\eta$) [Eq.~(\ref{mag_density})]; ALP mass ($m_a$) and
coupling ($\gag$); and the spectral parameters $C$ and $\Gamma$.  We
let the two spectral parameters vary from epoch to epoch, as they are
affected by the physical conditions at the $\g$-ray emission region at
different times, but keep the other four (two environmental, $\phi$
and $\eta$; and two ALP properties, $m_a$ and $\gag$) parameters fixed
in all epochs as they are not affected by the $\g$-ray emission
region.  We repeat this for three different electron density profiles:
$s=1$, 2 and 3 in Eq.~(\ref{mag_density}). The results of our fits are
shown in Fig.~\ref{fig:spectralfits} from the top to the bottom panels
for the electron density profile $s=3$ (a), 2 (b) and 1 (c),
respectively.

\FIGURE{\epsfig{file=fit_x3.eps, trim=0.cm 0.cm 0.cm 0.cm, clip,
    width=8.cm} \epsfig{file=fit_x2.eps, trim=0.cm 0.cm 0.cm 0.cm,
    clip, width=8.cm} \epsfig{file=fit_x1.eps, trim=0.cm 0.cm 0.cm
    0.cm, clip, width=8.cm}\caption{Fits to blazar 3C454.3 spectral
    data at 4 different epochs using ALP-photon mixing model in the
    blazar jet.  Each plot is for a particular electron density
    profile $n_e$ as a function of the jet radius $R$.  The parameters
    $m_a$ and $\gag$ are varied, but constrained to be the same at
    different epochs, together with the normalization and index of the
    production spectra, which are allowed to be different in different
    epochs.  (a) Profile $n_e \propto R^{-3}$.  Best-fit $m_a =
    2.5\cdot 10^{-7}$~eV and $\gag = 2.4\cdot 10^{-10}$~GeV$^{-1}$.
    (b) Profile $n_e \propto R^{-2}$.  Best-fit $m_a = 1.8\cdot
    10^{-7}$~eV and $\gag = 2.0\cdot 10^{-10}$~GeV$^{-1}$.  (c)
    Profile $n_e \propto R^{-1}$.  Best-fit $m_a = 1.1\cdot
    10^{-7}$~eV and $\gag = 3.3\cdot 10^{-10}$~GeV$^{-1}$. Note that
    we have not fitted the last data point in the Flare and Plateau
    epochs.}
\label{fig:spectralfits}}

The $\chi^2_{\rm min}$ values and the best-fit spectral parameters for
the case of $s=3$ (Fig.~\ref{fig:spectralfits} top panel) are: 24.5
(Flare, $C = 7.8\cdot 10^{-9}$, $\Gamma = 2.07$); 9.8 (Post-flare, $C
= 4.1\cdot 10^{-9}$, $\Gamma = 2.21$); 15.5 (Plateau, $C = 2.1\cdot
10^{-9}$, $\Gamma = 2.16$); and 11.0 (Quiet, $C = 7.8\cdot 10^{-10}$,
$\Gamma = 2.32$).  The best-fit environmental parameters are $\phi =
1.4\cdot 10^{-2}$, $\eta = 2.0\cdot 10^9$, and the best-fit ALP
parameters are $m_a = 2.5\cdot 10^{-7}$~eV, $\gag = 2.4\cdot
10^{-10}$~GeV$^{-1}$.

In the case of the $s=2$ electron density profile
(Fig.~\ref{fig:spectralfits} middle panel), the $\chi^2_{\rm min}$
values and the best-fit spectral parameters are: 23.3 (Flare, $C =
7.7\cdot 10^{-9}$, $\Gamma = 2.06$); 4.3 (Post-flare, $C = 4.1\cdot
10^{-9}$, $\Gamma = 2.20$); 17.0 (Plateau, $C = 2.1\cdot 10^{-9}$,
$\Gamma = 2.16$); and 10.2 (Quiet, $C = 7.8\cdot 10^{-10}$, $\Gamma =
2.32$).  The best-fit environmental parameters are $\phi = 1.6\cdot
10^{-2}$, $\eta = 2.0\cdot 10^9$, and the best-fit ALP parameters are
$m_a = 1.8\cdot 10^{-7}$~eV, $\gag = 2.0\cdot 10^{-10}$~GeV$^{-1}$.

In the case of $s=1$ electron density profile
(Fig.~\ref{fig:spectralfits} bottom panel) the $\chi^2_{\rm min}$
values and the best-fit spectral parameters are: 22.5 (Flare, $C =
8.0\cdot 10^{-9}$, $\Gamma = 2.09$); 6.2 (Post-flare, $C = 4.2\cdot
10^{-9}$, $\Gamma = 2.22$); 14.3 (Plateau, $C = 2.2\cdot 10^{-9}$,
$\Gamma = 2.18$); and 15.5 (Quiet, $C = 8.1\cdot 10^{-10}$, $\Gamma =
2.34$).  The best-fit environmental parameters are $\phi = 2.3\cdot
10^{-2}$, $\eta = 2.2\cdot 10^9$, and the best-fit ALP parameters are
$m_a = 1.1\cdot 10^{-7}$~eV, $\gag = 3.3\cdot 10^{-10}$~GeV$^{-1}$.

Note that the total $\chi^2$ values, summed over all of the four
epochs, are comparable for the 3 electron density profiles that we
have explored.  There is a slight preference for the $s=2$ profile.
This is compatible with the density profile in the jet deduced from
the reverberation measurement~\cite{Kaspi:1999}.  The $\sim 10$--20~mG
magnetic field and the $ \sim 2\cdot 10^9$~cm$^{-3}$ particle density
at a radius $R\sim 10^{18}$~cm, just at the edge or outside the BLR,
are also quite reasonable for the blazar jet.  The intrinsic spectrum
of $\g$ rays varies between $\Gamma \sim 2.1$--$2.3$ and are
compatible with the inverse Compton spectra by shock-accelerated
electrons, as generally thought to be the emission mechanism of $\g$
rays from blazars.

We can assess the significance of our results by adding up the
$\chi^2$ values for the four different epochs in each of the three
possible cases $s=1$, $2$ and $3$ and by comparing our results to the
$\chi^2$ resulting from a fit in which the probability of photon ALP
transition is absent and therefore there are only two free spectral
parameters $C$ and $\Gamma$. For the model which provides the best fit
to the data, that is, the $s=2$ model, the $\chi^2_{\rm{min,ALP}}$ in
the photon ALP mixing scenario is $54.8$ for $48$ spectral points and
$24$ free parameters, i.e. $24$ degrees of freedom (dof). If we
compute the $\chi^2$ without the photon ALP transition we obtain
$\chi^2_{\rm{min}}=412.9$, we have again $48$ spectral points to be
fitted with $8$ parameters, that is, $40$ degrees of freedom. The
difference in the fit for the two models, with and without photon ALP
mixing, is $\Delta \chi_{s=2}^2 = 358.1$ for $\Delta$(dof)$=16$. The
corresponding $p\simeq 0$ value is indicating that the ALP-photon
mixing model fit the data much better than with just a simple spectrum
characterized by Eq.~(\ref{spectrum}), that is, the simple model can
be rejected with a probability equal to $1$ for practical purposes.
For the $s=1$ and $s=3$ cases, $\Delta \chi^2_{s=1} = 354.4$ and
$\Delta \chi^2_{s=3} = 352.1$ respectively and therefore $p\simeq 0$
also for these two cases.

The results quoted above have been obtained for the $\xi= \pi/4$,
angle which describes the configuration of the transverse magnetic
field $B_T$. Similar results could be obtained for different magnetic
field configurations, although with slightly different best-fit values
for the parameters describing the model.

Figure~\ref{fig:contourplots} shows $68\%$ (yellow) and $95\%$ (green)
confidence-level regions in the $m_a$-$\gag$ parameter space for our
ALP-photon mixing model for the $\g$-ray spectral data of the blazar
3C454.3.  For these plots ($s=3$, $2$ and $1$ for the left, middle and
right panel, respectively) we have kept marginalized over the spectral
parameters $C$ and $\Gamma$ as well as over the environmental
parameters $\phi$ and $\eta$, i.e. the four parameters which are
common to all epochs.  The best-fit points in the $m_a$-$\gag$
parameter space are denoted with a ``$*$'' and these values are in
agreement with the fits obtained in Fig.~\ref{fig:spectralfits}.  Note
that the allowed values for the ALP mass and its coupling depend
rather weakly on the particle density profile in the blazar jet and
are constrained in small ranges, $m_a \sim (0.8-2.5)\cdot 10^{-7}$~eV
and $\gag \sim (1.2-3.3)\cdot 10^{-10}$~GeV.

\FIGURE{\epsfig{file=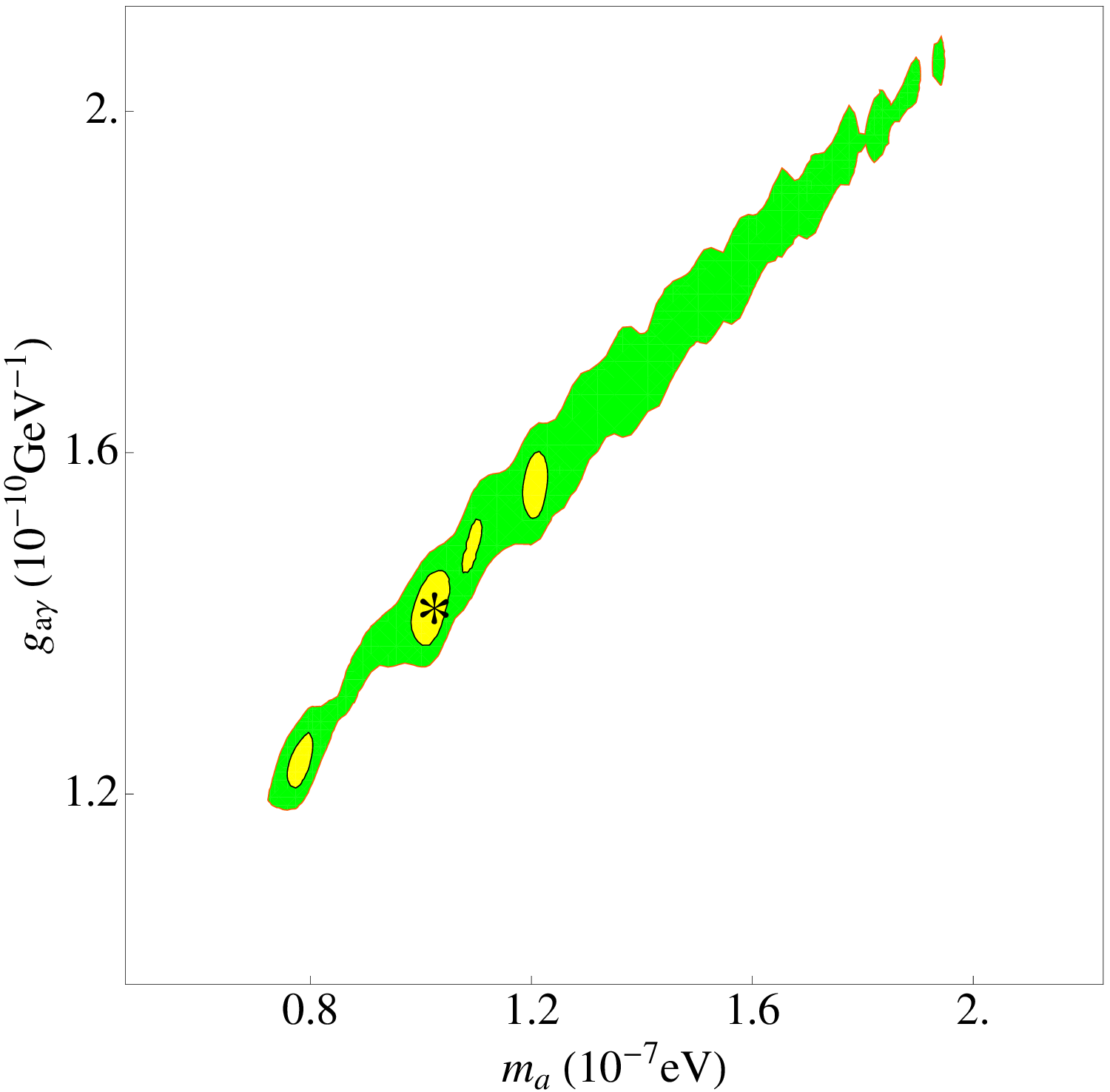, trim=0.cm 0.cm 0.cm 0.cm, clip,
    height=4.8cm} \epsfig{file=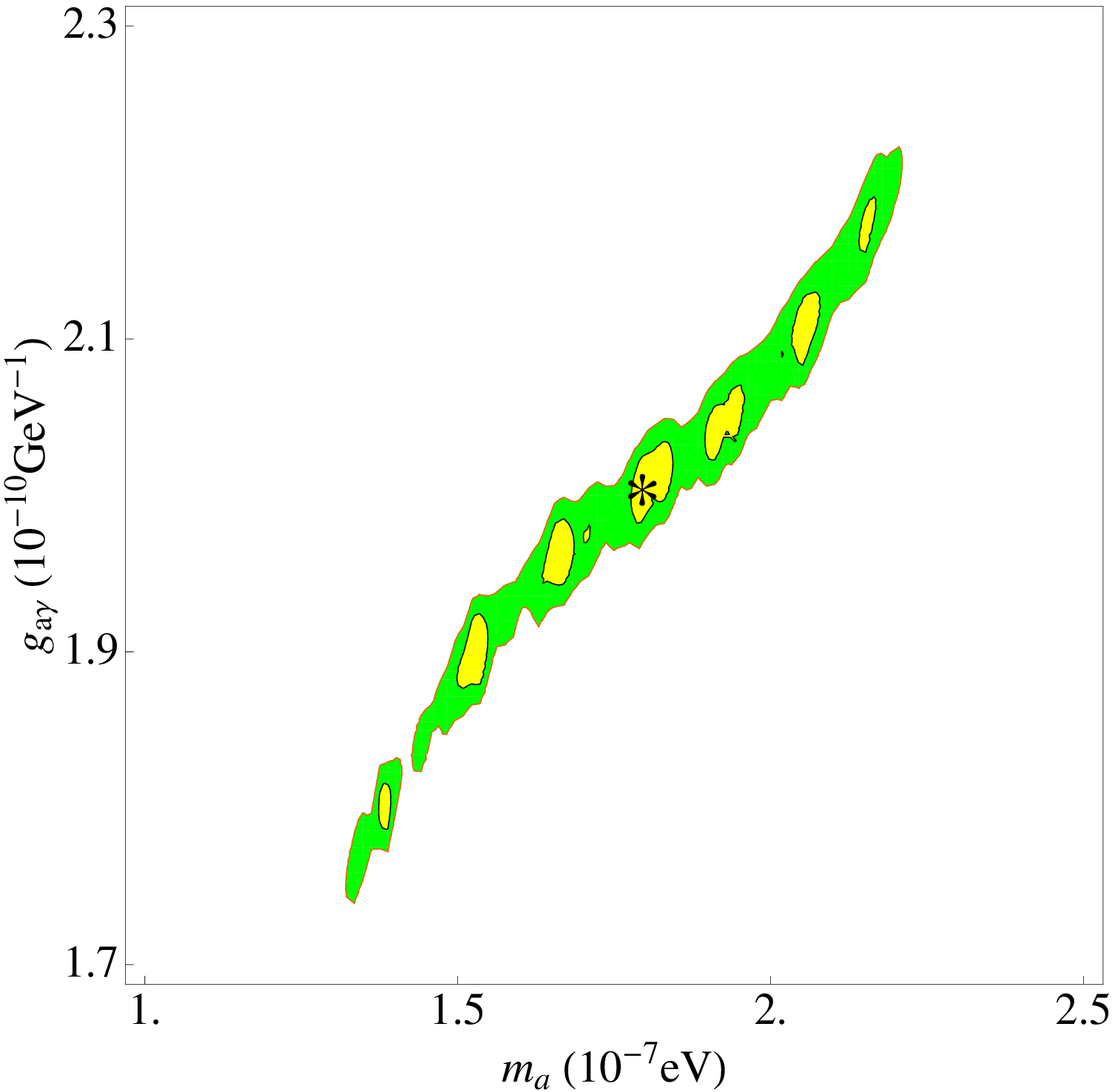, trim=0.cm 0.cm 0.cm
    0.cm, clip, height=4.8cm} \epsfig{file=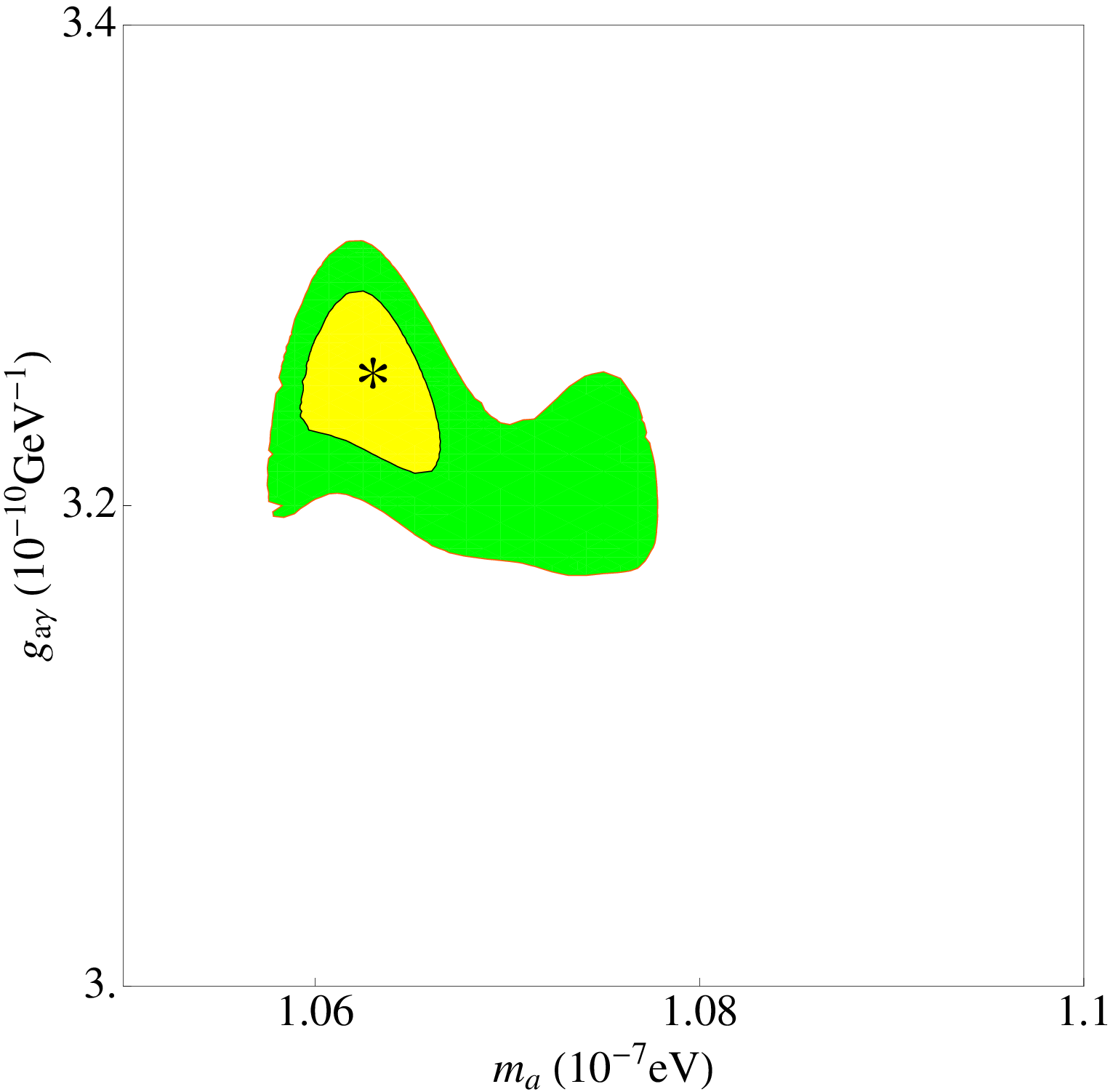, trim=0.cm
    0.cm 0.cm 0.cm, clip, height=4.8cm} \caption{Contour plots of the
    $68\%$ and $95\%$ CL regions for the ALP mass and coupling in case
    of the electron density profiles in the jet (from left to right)
    $n_e\propto R^{-3}$, $n_e\propto R^{-2}$ and $n_e\propto R^{-1}$.
    The parameters of the production spectra for each epoch found in
    Fig.~\ref{fig:spectralfits} are kept fixed.  The position of the
    symbol ``$*$'' denotes the best-fit point in the $m_a$-$\gag$
    plane.}
\label{fig:contourplots}}

\section{Consistency check with PKS1222+216 data}

{\em Fermi}-LAT and MAGIC Cherenkov telescope made overlapping
observations of PKS1222+216 on 2010 June 17
\cite{Tanaka:2011gd,Aleksic:2011hr}.  During this time {\em Fermi}-LAT
measured $\gamma$-ray emission in the $\sim 0.3$--1 GeV range and
MAGIC measured VHE $\gamma$-ray emission in the $\sim 70$--400 GeV
range.  These data are showin in Fig.~\ref{fig:pks1222}. VHE $\gamma$
rays are subject to absorption due to $\gamma\gamma\to e^+e^-$ pair
production with UV-infrared photons of the extragalactic background
light (EBL).  Figure~\ref{fig:pks1222} also shows the absorption
corrected VHE ``deabsorbed'' spectrum as reported in
Ref.~\cite{Aleksic:2011hr} which used the EBL model in
Ref.~\cite{Dominguez:2010bv}.  The EBL model in
Refs.~\cite{Razzaque:2008te, Finke:2009xi} also gives very similar
deabsorbed VHE spectrum.

\FIGURE{\epsfig{file=PKS1222_allR.eps, trim=0.cm 0.cm 0.cm 0.cm, clip,
    width=10.cm}
  \caption{Blazar PKS1222+216 spectral data (filled symbols) from {\em
      Fermi}-LAT and MAGIC Cherenkov telescope. MAGIC-deabsorbed data
    points (empty symbols) correspond to EBL absorption corrected
    spectrum.  The ALP model fits, for different electron density
    profiles $n_e \propto R^{-s}$, are performed using the {\em
      Fermi}-LAT and MAGIC deabsorbed data, while keeping $m_a$ and
    $\gag$ values fixed for each $s$ values as obtained in
    Fig.~\ref{fig:spectralfits}.}
\label{fig:pks1222}}

Note that $\gtrsim 10$ GeV $\gamma$ rays, if the emission region is
well below the BLR, are also subject to absorption by BLR
photons~\cite{Poutanen:2010he}.  It was suggested in
Ref.~\cite{Tavecchio:2012um} that ALP-photon mixing in the BLR reduces
the $\gamma\gamma$ optical depth as ALPs are not subject to
absorption.  The authors in Ref.~\cite{Tavecchio:2012um} also fitted
the high-energy and VHE spectra of PKS1222+216 using emissions,
respectively, from two blobs and by assuming ALP-photon mixing in the
BLR.  The resulting ALP parameters are $m_a \sim 10^{-10}$~eV and
$\gag \sim 1.4\times 10^{-11}$. These are much different than our
values from fitting multi-epoch 3C454.3 data, assuming that the
$\gamma$-ray emission region is beyond BLR, in Sec.\ 3.

To verify that the ALP parameter values we have obtained are
consistent with PKS1222+216 data, we have fitted the {\em Fermi}-LAT
and MAGIC deabsorbed spectra assuming our simpler one-zone emission
(in the form of a single power-law) region beyond the BLR.  We have
kept the $m_a$ and $\gag$ values fixed for $s=1$, 2 and 3 as in
Fig.~\ref{fig:spectralfits} but varied the environmental ($\phi$,
$\eta$) and spectral ($C$, $\Gamma$) parameters.  The resulting fits
with best-fit parameter values are shown in Fig.~\ref{fig:pks1222}.

The $\chi^2_{\rm min}$ values and the best-fit environmental and
spectral parameters for the case of $s=3$ are: 1.3, $\phi = 6\cdot
10^{-3}$, $\eta = 10^{8}$, $C = 1.4\cdot 10^{-9}$ and $\Gamma = 2.17$.
In case of $s=2$, these values are: 1.5, $\phi = 6\cdot 10^{-3}$,
$\eta = 10^{7}$, $C = 1.4\cdot 10^{-9}$ and $\Gamma = 2.19$.  Finally
in case of $s=1$, these values are: 6.1, $\phi = 6\cdot 10^{-3}$,
$\eta = 2.4\cdot 10^{9}$, $C = 1.4\cdot 10^{-9}$ and $\Gamma = 2.19$.
A single power-law fit without ALP mixing, on the hand, gives
$\chi^2_{\rm min} = 3.4$.  Except for the $s=1$ case, the
$\Delta\chi^2_{\rm min}$ for 2 dof difference are quite good.  The
parameter values are also quite reasonable for blazars.

\section{Discussion and conclusions}

The ALP mass, $m_a \sim (1-3)\cdot 10^{-7}$~eV, and coupling, $\gag
\sim (1-3)\cdot 10^{-10}$~GeV$^{-1}$, that we have obtained from
fitting spectral data of the well-studied blazar 3C454.3 in the GeV
energy range, are just at the border of the exclusion zone in the
$m_a$-$\gag$ parameter space from the Cern Axion Solar Telescope
(CAST) experiment~\cite{Arik:2011rx}, as well as close to the region
excluded by axion energy losses induced in massive
stars~\cite{Friedland:2012hj} $g_{a\gamma} \lesssim (0.8-1) \cdot
10^{-10}$ GeV$^{-1}$.  Note, however, that significant astrophysical
model uncertainties might be present when deriving the former bound.
Our results are also consistent with the recent exclusion region in
the $m_a$-$\gag$ parameter space from the laboratory experiment of
``Light Shining through a Wall'' by the ALPS
Collaboration~\cite{Ehret:2010mh}.  A new generation of axion
helioscopes~\cite{Irastorza:2011gs} will be able to probe the
parameter space which includes our best-fit $m_a$ and $\gag$ values.

Interestingly the ALP mass that we have found in this work is very
similar to the mass required to produce a spectral feature observed in
the GRB data (in about 15\% cases) through the similar ALP-photon
mixing mechanism, but in a much shorter scale, $\sim
10^{13}$--$10^{14}$~cm, jet with a much stronger, $\sim
10^4$--$10^5$~G, magnetic field~\cite{Mena:2011xj}.  On the other
hand, our ALP mass is 3 orders of magnitude larger than the $\sim
10^{-10}$~eV mass suggested in Ref.~\cite{Tavecchio:2012um} from
modeling of $> 10$~GeV $\g$-ray data from PKS1222+216, assuming the
$\g$-ray emission region is well below the BLR, contrary to our model.
Moreover, ALP-photon mixing scenario was invoked in
Ref.~\cite{Tavecchio:2012um} to alleviate strong $\g\g$ absorption of
VHE photons expected in the BLR rather than the direct flux
suppression effect that we have explored.  Using the same PKS1222+216
data we have obtained good fits for our ALP-photon mixing scenario for
$m_a \sim 10^{-7}$ eV.

Very light, $\lesssim 10^{-10}$~eV, ALP mass is favored for mixing
with VHE photons in the intergalactic magnetic field (IGMF) of
megaparsec scale coherence length.  Indeed for $\sim 1$~nG IGMF, $\sim
10^{-10}$~eV ALPs mix strongly with $\omega_L \sim 10\,
(m_a/10^{-10}~{\rm eV})^2 (B_T/{\rm nG})^{-1}$~GeV or higher energy
photons, for $\gag \sim 10^{-10}$~GeV$^{-1}$ and for typical particle
densities of $n_e\sim 10^{-7}$~cm$^{-3}$ in the intergalactic media.
We calculate below the effect of IGMF on the $\sim 10^{-7}$~eV ALP
mass that we have found.

The ALP-photon conversion probability in the IGMF can be written as
(see, e.g., Ref.~\cite{Mena:2011xj})
\begin{equation}
P_{a \gamma} =\sin^2 2 \theta
\sin^2 \left( \frac{\Delta_{\rm osc} \, L}{2} \right),
\label{eq:probosc}
\end{equation}
where $L\sim 1$~Mpc and the oscillation wave number and the mixing
angle are given by ${\Delta}_{\rm osc} = \sqrt{(\Delta_a -
  \Delta_{\parallel})^2 + 4 \Delta_{a\g}^2}$ and $\theta = (1/2)
\arctan [2 \Delta_{a \gamma}/(\Delta_{\parallel}-\Delta_a)]$,
respectively.  For our case of $m_a\sim 10^{-7}$~eV, $\Delta_{\rm osc}
\approx |\Delta_a|$, where $\Delta_a$ is given in
Eq.~(\ref{matrix_elements}), which is independent of any magnetic
field.  Since $L\gg \Delta_{\rm osc}$, the oscillation term in
Eq.~(\ref{eq:probosc}) averages out.  By requiring that the amplitude
of oscillations be at least $1/2$ for ALPs from blazar 3C454.3 to be
converted back to photons, we get $2\theta \approx \arctan
(2\Delta_{a\g}/|\Delta_a|) \gtrsim \arcsin(1/\sqrt{2})= \pi/4$.  Thus
the ALPs will convert back to photons for
\be 
B_{\rm IGMF} \gtrsim 8.4\cdot 10^{-3} \left(\frac{E}{\rm
    GeV}\right)^{-1} \left(\frac{\gag}{10^{-10}~{\rm
      GeV}^{-1}}\right)^{-1} \left(\frac{m_a}{10^{-7}~{\rm
      eV}}\right)^{2} ~{\rm G}.
\label{IGMF_limit}
\ee
This is already above any reasonable estimate of the IGMF, as well as
the magnetic field in the clusters of galaxies.  A similar analysis
can be performed for the Galactic $\mu$G magnetic field.  Thus the
spectral feature in the $\sim 1$--10~GeV range for the blazar 3C454.3,
due to conversions of photons to ALPs in the kpc scale jet, remains
unchanged.

Explanations of the GeV spectral breaks seen in {\em Fermi-}LAT
detected blazars without invoking new physics have been attempted
earlier.  These include $\g\g$ absorption of GeV photons by the Lyman
line and continuum radiation from He {\small II} in the
BLR~\cite{Poutanen:2010he}, and two component GeV emission from
Compton scattering of accretion disc photons and BLR
photons~\cite{Finke:2010rg}.  While such scenarios can be responsible
for the observed GeV spectral breaks, we note that in both scenarios
the $\g$-ray emission region is below or within the BLR, contrary to
our assumption.  A future systematic study of GeV spectral breaks in
all blazars will shed further light on the hints of ALP-photon mixing
that we have found for the blazar 3C454.3 and can provide clues to
distinguish between a conventional and exotic explanation of the
breaks in the blazar spectra.

\section{Acknowledgments}

We thank Charles Dermer, Justin Finke, Benoit Lott, Andreas Ringwald
and Pierre Sikivie for discussion and comments.  O.M. is supported by
the Consolider Ingenio project CSD2007-00060, by PROMETEO/2009/116, by
the Spanish Ministry Science project FPA2011-29678 and by the ITN
Invisibles PITN-GA-2011-289442.


\begin{thebibliography}{srt}
\def\bitm{\bibitem}

\bibitem{Sikivie:1983ip} 
 P.~Sikivie,
  Phys.\ Rev.\ Lett.\  {\bf 51}, 1415 (1983)
  [Erratum-ibid.\  {\bf 52}, 695 (1984)].

\bibitem{Sikivie:1985yu} 
  P.~Sikivie,
  Phys.\ Rev.\ D {\bf 32}, 2988 (1985)
  [Erratum-ibid.\ D {\bf 36}, 974 (1987)].

\bibitem{Raffelt:1987im} 
  G.~Raffelt and L.~Stodolsky,
  Phys.\ Rev.\ D {\bf 37}, 1237 (1988).

\bibitem{Ringwald:2012hr} 
  A.~Ringwald,
  Phys.\ Dark Univ.\  {\bf 1}, 116 (2012)
  [arXiv:1210.5081 [hep-ph]].

\bibitem{Peccei:1977hh} 
  R.~D.~Peccei and H.~R.~Quinn,
  Phys.\ Rev.\ Lett.\  {\bf 38}, 1440 (1977).

\bibitem{Arvanitaki:2009fg} 
  A.~Arvanitaki, S.~Dimopoulos, S.~Dubovsky, N.~Kaloper and J.~March-Russell,
  Phys.\ Rev.\ D {\bf 81}, 123530 (2010)
  [arXiv:0905.4720 [hep-th]].

\bibitem{Acharya:2010zx} 
  B.~S.~Acharya, K.~Bobkov and P.~Kumar,
  JHEP {\bf 1011}, 105 (2010)
  [arXiv:1004.5138 [hep-th]].

\bibitem{Cicoli:2012sz} 
  M.~Cicoli, M.~Goodsell, A.~Ringwald, M.~Goodsell and A.~Ringwald,
  JHEP {\bf 1210}, 146 (2012)
  [arXiv:1206.0819 [hep-th]].

\bibitem{Ringwald:2012cu} 
  A.~Ringwald,
  arXiv:1209.2299 [hep-ph].

\bibitem{Arik:2011rx} 
  S.~Aune {\it et al.}  [CAST Collaboration],
  Phys.\ Rev.\ Lett.\  {\bf 107}, 261302 (2011)
  [arXiv:1106.3919 [hep-ex]].

\bibitem{Friedland:2012hj} 
  A.~Friedland, M.~Giannotti and M.~Wise,
  Phys.\ Rev.\ Lett.\  {\bf 110}, 061101 (2013)
  [arXiv:1210.1271 [hep-ph]].

\bibitem{Raffelt:1987yu}
  G.~G.~Raffelt and D.~S.~P.~Dearborn,
  Phys.\ Rev.\ D {\bf 36}, 2211 (1987).

\bibitem{Bahre:2013ywa} 
  R.~Bähre, B.~Döbrich, J.~Dreyling-Eschweiler, S.~Ghazaryan, R.~Hodajerdi, D.~Horns, F.~Januschek and E.~-A.~Knabbe {\it et al.},
  arXiv:1302.5647 [physics.ins-det].

\bibitem{Hooper:2007bq} 
  D.~Hooper and P.~D.~Serpico,
  Phys.\ Rev.\ Lett.\  {\bf 99}, 231102 (2007)
  [arXiv:0706.3203 [hep-ph]].

\bibitem{Tavecchio:2012um} 
  F.~Tavecchio, M.~Roncadelli, G.~Galanti and G.~Bonnoli,
  Phys.\ Rev.\ D {\bf 86}, 085036 (2012)
  [arXiv:1202.6529 [astro-ph.HE]].

\bibitem{Mena:2011xj} 
  O.~Mena, S.~Razzaque and F.~Villaescusa-Navarro,
  JCAP {\bf 1102}, 030 (2011)
  [arXiv:1101.1903 [astro-ph.HE]].

\bibitem{SanchezConde:2009wu} 
  M.~A.~Sanchez-Conde, D.~Paneque, E.~Bloom, F.~Prada and A.~Dominguez,
  Phys.\ Rev.\ D {\bf 79}, 123511 (2009)
  [arXiv:0905.3270 [astro-ph.CO]].

\bibitem{Bassan:2010ya} 
  N.~Bassan, A.~Mirizzi and M.~Roncadelli,
  JCAP {\bf 1005}, 010 (2010)
  [arXiv:1001.5267 [astro-ph.HE]].

\bibitem{Horns:2012kw} 
  D.~Horns, L.~Maccione, M.~Meyer, A.~Mirizzi, D.~Montanino and M.~Roncadelli,
  Phys.\ Rev.\ D {\bf 86}, 075024 (2012)
  [arXiv:1207.0776 [astro-ph.HE]].

\bibitem{Simet:2007sa} 
  M.~Simet, D.~Hooper and P.~D.~Serpico,
  Phys.\ Rev.\ D {\bf 77}, 063001 (2008)
  [arXiv:0712.2825 [astro-ph]].

\bibitem{Abdo:2011} 
  A.~A.~Abdo, M.~Ackermann, M.~Ajello, A.~Allafort, L.~Baldini, J.~Ballet, G.~Barbiellini and D.~Bastieri {\it et al.},
  Astrophys.\ J.\ Lett.\ {\bf 733}, L26 (2011)

\bibitem{Tanaka:2011gd} 
  Y.~T.~Tanaka, L.~Stawarz, D.~J.~Thompson, F.~D'Ammando, S.~J.~Fegan, B.~Lott, D.~L.~Wood and C.~C.~Cheung {\it et al.},
  Astrophys.\ J.\  {\bf 733}, 19 (2011)
  [arXiv:1101.5339 [astro-ph.HE]].


\bibitem{Aleksic:2011hr} 
  J.~Aleksic {\it et al.}  [MAGIC Collaboration],
  Astrophys.\ J.\  {\bf 730}, L8 (2011)
  [arXiv:1101.4645 [astro-ph.HE]].


\bibitem{Aharonian:2007ig} 
  F.~Aharonian,
  Astrophys.\ J.\  {\bf 664}, L71 (2007)
  [arXiv:0706.0797 [astro-ph]].

\bibitem{Albert:2007zd} 
  J.~Albert, E.~Aliu, H.~Anderhub, P.~Antoranz, A.~Armada, C.~Baixeras, J.~A.~Barrio and H.~Bartko {\it et al.},
  Astrophys.\ J.\  {\bf 669}, 862 (2007)
  [astro-ph/0702008].

\bibitem{Marscher:2008aa} 
  A.~P.~Marscher, S.~G.~Jorstad, F.~D.~D'Arcangelo, P.~S.~Smith, G.~G.~Williams, V.~M.~Larionov, H.~Oh and A.~R.~Olmstead {\it et al.},
  Nature {\bf 452}, 966 (2008).

\bibitem{Giannios:2009kh} 
  D.~Giannios, D.~A.~Uzdensky and M.~C.~Begelman,
  Mon.\ Not.\ Royal Astron.\ Soc.\ Lett.\ {\bf 395}, L29 (2009) 
  [arXiv:0901.1877 [astro-ph.HE]].

\bibitem{Begelman:1984mw} 
  M.~C.~Begelman, R.~D.~Blandford and M.~J.~Rees,
  Rev.\ Mod.\ Phys.\  {\bf 56}, 255 (1984).

\bibitem{Rees:1987} 
  M.~J.~Rees,
  Mon.\ Not.\ Roy.\ Astron.\ Soc.\  {\bf 228}, 47 (1987).

\bibitem{Kaspi:1999} 
  S.~Kaspi and H.~Netzer,
  Astrophys.\ J.\ {\bf 524}, 71 (1999)

\bibitem{Dominguez:2010bv} 
  A.~Dominguez, J.~R.~Primack, D.~J.~Rosario,
  F.~Prada, R.~C.~Gilmore, S.~M.~Faber, D.~C.~Koo and R.~S.~Somerville
  {\it et al.},
  Mon.\ Not.\ Royal Astron.\ Soc.\ {\bf 410}, 2556 (2011)
  [arXiv:1007.1459 [astro-ph.CO]].

\bibitem{Razzaque:2008te} 
  S.~Razzaque, C.~D.~Dermer and J.~D.~Finke,
  Astrophys.\ J.\  {\bf 697}, 483 (2009)
  [arXiv:0807.4294 [astro-ph]].

\bibitem{Finke:2009xi} 
  J.~D.~Finke, S.~Razzaque and C.~D.~Dermer,
  Astrophys.\ J.\  {\bf 712}, 238 (2010)
  [arXiv:0905.1115 [astro-ph.HE]].

\bibitem{De Angelis:2007yu} 
  A.~De Angelis, O.~Mansutti and M.~Roncadelli,
  Phys.\ Lett.\ B {\bf 659}, 847 (2008)
  [arXiv:0707.2695 [astro-ph]].

\bibitem{Ehret:2010mh} 
  K.~Ehret, M.~Frede, S.~Ghazaryan, M.~Hildebrandt, E.~-A.~Knabbe, D.~Kracht, A.~Lindner and J.~List {\it et al.},
  Phys.\ Lett.\ B {\bf 689}, 149 (2010)
  [arXiv:1004.1313 [hep-ex]].

\bibitem{Irastorza:2011gs} 
  I.~G.~Irastorza, F.~T.~Avignone, S.~Caspi, J.~M.~Carmona, T.~Dafni, M.~Davenport, A.~Dudarev and G.~Fanourakis {\it et al.},
  JCAP {\bf 1106}, 013 (2011)
  [arXiv:1103.5334 [hep-ex]].

\bibitem{Poutanen:2010he} 
  J.~Poutanen and B.~Stern,
  Astrophys.\ J.\  {\bf 717}, L118 (2010)
  [arXiv:1005.3792 [astro-ph.HE]].

\bibitem{Finke:2010rg} 
  J.~D.~Finke and C.~D.~Dermer,
  Astrophys.\ J.\  {\bf 714}, L303 (2010)
  [arXiv:1004.1418 [astro-ph.HE]].


\end{thebibliography}
\end{document}